\documentclass[10pt,sigconf,final]{IEEEtran}
\IEEEoverridecommandlockouts
\usepackage{amsmath,amssymb,amsfonts}
\usepackage{algorithmic}
\usepackage{graphicx}
\usepackage{hyperref}

\usepackage{Alegreya}
\RequirePackage{enumitem}
\RequirePackage[disable]{todonotes}
\usepackage{listing}
\RequirePackage{tcolorbox}
\tcbuselibrary{skins,listings,breakable}

\newtcblisting{boxyamlnote}{
  sharpish corners, %
  boxrule = 0pt,
  toprule = 1.5pt, %
  enhanced,
  drop fuzzy shadow = white,
  fontupper=\sffamily,
  listing only,
  listing options={
    numbers=left
  }
}

\newtcblisting{boxshellnote}{
  sharpish corners, %
  boxrule = 0pt,
  toprule = 1.5pt, %
  enhanced,
  drop fuzzy shadow = white,
  fontupper=\sffamily,
  listing only,
}

\newtcblisting{boxtablenote}{
  sharpish corners, %
  boxrule = 0pt,
  toprule = 1.5pt, %
  enhanced,
  drop fuzzy shadow = white,
  fontupper=\sffamily,
  listing only,
}

\hypersetup{
  colorlinks=true,
  linkcolor=blue,
  citecolor=blue,
  urlcolor=blue
}

\usepackage{orcidlink}
\def\BibTeX{{\rm B\kern-.05em{\sc i\kern-.025em b}\kern-.08em
T\kern-.1667em\lower.7ex\hbox{E}\kern-.125emX}}
\usepackage{amsthm}
\usepackage[
sortcites = true,
sorting=none,
backend=biber
]{biblatex}
\usepackage{siunitx}
\usepackage{xspace}
\usepackage{booktabs}

\usepackage[acronym]{glossaries}

\newacronym{ci}{CI}{Continuous Integration}
\newacronym{cb}{CB}{Continuous Benchmarking}
\newacronym{cx}{CX}{Continuous eXperimentation}
\newacronym{hpc}{HPC}{High Performance Computing}
\newglossaryentry{exacb}{%
  name=exaCB,
  description={Tool for Continuous Benchmarking on an Exascale level}
}
\newacronym{cicd}{CI/CD}{Continuous Integration and Delivery}

\usepackage{tgpagella}
\usepackage[svgnames,x11names]{xcolor}
\usepackage{hyperref}

\colorlet{central}{DodgerBlue2}
\colorlet{couple}{MediumPurple2}
\colorlet{centralyes}{central!80!white}
\colorlet{centralno}{central!80!black}
\colorlet{coupleyes}{couple!80!white}
\colorlet{coupleno}{couple!80!black}

\usepackage{tikz}
\newcommand{\circled}[1]{%
  \tikz[baseline=(char.base)]{
    \node[shape=circle, inner sep=1pt, font=\footnotesize, fill=lightgray, text=white] (char) {#1};
  }%
}

\newcommand{\exacb}{\gls{exacb}}
\glsdisablehyper%

\usepackage{xcolor}

\PassOptionsToPackage{most}{tcolorbox}

\definecolor{DelBg}{RGB}{255,230,230}
\definecolor{DelStroke}{RGB}{180,0,0}

\newtcolorbox{delblock}{
  enhanced,
  breakable,
  colback=red!6,
  colframe=red!60!black,
  boxrule=0.4pt,
  borderline west={2pt}{0pt}{red!70!black},
  left=6pt,right=4pt,top=3pt,bottom=3pt,
  overlay={%
    \draw[DelStroke, line width=0.8pt, opacity=0.45]
      (frame.north west) -- (frame.south east);
    \draw[DelStroke, line width=0.8pt, opacity=0.45]
      (frame.south west) -- (frame.north east);
  },
}

\newtcolorbox{addblock}{
  enhanced,
  breakable,
  colback=green!6,
  colframe=green!50!black,
  boxrule=0.4pt,
  borderline west={2pt}{0pt}{green!60!black},
  left=6pt,right=4pt,top=3pt,bottom=3pt,
}

\newtcbox{\delete}{ %
  on line,
  colback=red!6,
  colframe=red!60!black,
  boxrule=0.4pt,
  left=1pt,right=1pt,top=1pt,bottom=1pt,
  boxsep=0pt,
}

\newtcbox{\add}{ %
  on line,
  colback=green!6,
  colframe=green!50!black,
  boxrule=0.4pt,
  left=1pt,right=1pt,top=1pt,bottom=1pt,
  boxsep=0pt,
}

\usepackage{listings}
\usepackage{xcolor} %

\begin{document}
\pagestyle{plain} %

\title{exaCB: Reproducible Continuous Benchmark Collections at Scale Leveraging an Incremental Approach}

\author{\IEEEauthorblockN{Jayesh Badwaik\orcidlink{0000-0002-5252-8179}}
  \ \
  \and
  \IEEEauthorblockN{Mathis Bode\orcidlink{0000-0001-9922-9742}} \ \
  \and
  \IEEEauthorblockN{Michal Rajski\orcidlink{0009-0001-4683-1841}} \ \
  \and
  \IEEEauthorblockN{Andreas Herten\,\orcidlink{0000-0002-7150-2505}}\\
  \IEEEauthorblockA{\textit{Jülich Supercomputing Centre} \\
    \textit{Forschungszentrum Jülich}\\
  Jülich, Germany}
}
\maketitle
\todo[inline]{back end, front end, but then back-end processes}

\begin{abstract}
The increasing heterogeneity of high-performance computing (HPC)
systems and the transition to exascale architectures require
systematic and reproducible performance evaluation across diverse
workloads. While continuous integration (CI) ensures functional
correctness in software engineering, performance and energy efficiency
in HPC are typically evaluated outside CI workflows, motivating
continuous benchmarking (CB) as a complementary approach. Integrating
benchmarking into CI workflows enables reproducible evaluation, early
detection of regressions, and continuous validation throughout the
software development lifecycle.

We present \exacb{}, a framework for continuous benchmarking developed
in the context of the JUPITER exascale system. \exacb{} enables
application teams to integrate benchmarking into their workflows while
supporting large-scale, system-wide studies through reusable CI/CD
components, established harnesses, and a shared reporting protocol.
The framework supports incremental adoption, allowing benchmarks to be
onboarded easily and to evolve from basic runnability to more advanced
instrumentation and reproducibility. The approach is demonstrated in
JUREAP, the early-access program for JUPITER, where \exacb{} enabled
continuous benchmarking of over 70 applications at varying maturity
levels, supporting cross-application analysis, performance tracking,
and energy-aware studies. These results illustrate the practicality 
using \exacb{} for continuous benchmarking for exascale HPC systems 
across large, diverse collections of scientific applications.
\end{abstract}

\section{Introduction}
\label{sec:introduction}

High-performance computing (HPC) systems increasingly combine heterogeneous
architectures with complex, tightly coupled software stacks to deliver optimal 
performance for diverse user workloads~\cite{gamblin2015,allen2020}. To ensure that systems 
and software perform as expected, reliable and continuous performance evaluation across a broad range of 
applications is essential. Such evaluation is critical during system procurement and acceptance, early access and
bring-up phases, and throughout regular operation to ensure consistent performance over time.

This manuscript presents \exacb{}, a framework for continuous benchmarking of scientific workloads on HPC systems, utilizing established software engineering practices to aid administrators, support staff, technology explorers, and application users in ensuring optimal performance.

\subsection{Challenges of HPC Workloads}

Traditionally, system performance has often been addressed by using a set of well-known benchmarks, assuming that the results would generalize to a wide range
of applications. These benchmarks would often be carefully designed and maintained by the system
operators themselves, potentially independently of development of the upstream software, with
only occasional updates during major events such as system upgrades or procurement of new systems.
In an era where systems were relatively homogeneous, and applications were relatively simple, this
approach was often sufficient to provide a reasonable estimate of the system performance for a wide
range of applications.

However, modern HPC hardware is increasingly heterogeneous~\cite{carpenter2022}, with a variety of architectures
becoming available, each with its own strengths and weaknesses. Furthermore, the applications
themselves are becoming increasingly complex, often involving multiple components and large number
of dependencies that interact in non-trivial ways. In such an environment, the traditional approach
of using a small set of benchmarks is no longer sufficient to capture the performance
characteristics of the full set of systems that a computing center may offer.

At the same time, maintaining a large and diverse set of benchmarks that is broad enough to cover
the full spectrum of systems and applications is a daunting task, and often infeasible without
sustained support from the upstream software community. As a result, the traditional model of
maintaining benchmarks independently of upstream development, and in a largely static manner, is no
longer sufficient. Instead, benchmarking must be integrated more directly into the software
development process so that application developers, system administrators, and vendors can rely on
shared code paths to enable consistent, reproducible performance evaluation across systems.

The complexity of the modern HPC ecosystem necessitates a shift in how benchmarking is approached in
itself as well. It is often no longer sufficient to consider each benchmark independently, but
rather as a part of a larger ecosystem of benchmarks that need to be maintained and executed in a
collaborative manner. A change in a system component may have divergent effects on different
benchmarks, and only by considering the full spectrum of benchmarks can system operators and
computing centers make informed optimization decisions.

Also the benchmarking ecosystem introduces its own challenges. The diversity of
hardware architectures means that not all benchmarks are ready to run on all systems at all times \cite{koskela2023};
moreover, components throughout the benchmarking stack evolve at different paces across systems,
leading to varying levels of benchmark readiness. A systematic benchmarking approach must therefore be flexible
enough to accommodate this diversity, not only in methodologies but also in levels of readiness,
while enabling benchmarks to continuously mature toward a high level of readiness.

Finally, in the era of exascale computing, energy efficiency becomes a first-class concern \cite{kogge2008,heldens2021};
benchmarks must therefore be not only runnable, but also readily instrumentable and amenable to
performance, power, and energy analysis and optimization. Such instrumentation should be supported
from the outset and often includes performance-based measurements which, even when not directly
required, facilitate analysis and significantly simplify sharing and reuse.

These challenges require a conceptual change in how HPC workloads are executed, establishing new methodologies for reproducible workflows to focus on close collaboration between the application developers, support engineers, and system administrators.

\subsection{Collaborative benchmarking}

\begin{figure*}[tb]
  \centering
  \includegraphics[width=\linewidth]{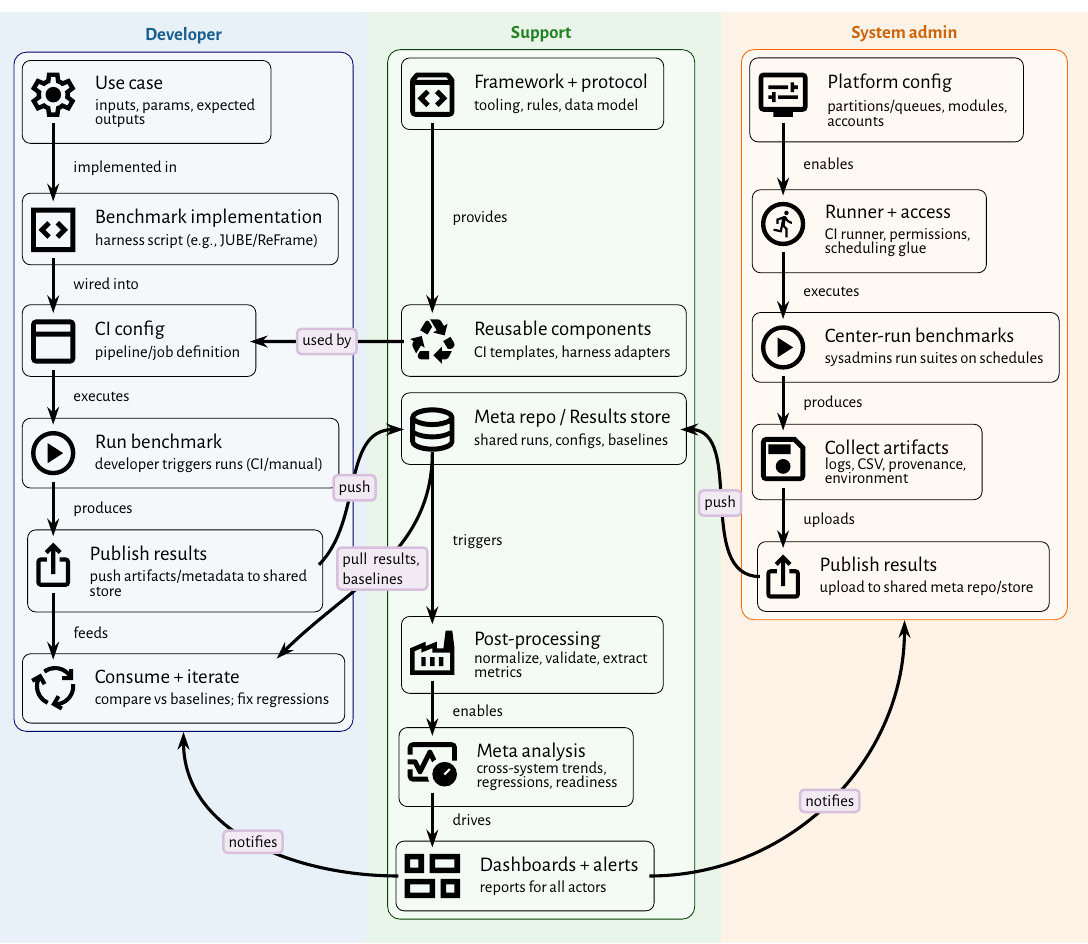}
  \caption{Collaborative benchmarking methodology for modern HPC systems}
  \label{fig:collaborative_benchmarking}
\end{figure*}

The general concept of collaborative benchmarking is schematically represented in \autoref{fig:collaborative_benchmarking}. Here, benchmarking is inherently collaborative, relying on a direct connection between developers, support engineers, and system administrators. Each of these actors has their own set of tasks that together form a pipeline, interleaved with the exchange of services. Reusable CI components, provided by support entities, enable developers to run benchmarks in a standardized way, while data produced by both developers and administrators is collected into a common meta-repository or result store. This data can then be post-processed and meta-analyzed to extract insights about application and system performance. The resulting information drives the reporting component, which on one hand allows developers to iterate and investigate the performance of their application, while on the other hand provides administrators with insights about the system they operate. Through this workflow, the benchmarking process is automated and ensures a rigorous and reproducible execution of the application in continuous fashion.

\subsection{Continuous Benchmarking}

In this work, we present \exacb, a framework for continuous benchmarking on HPC systems that
addresses the challenges outlined above, up to exascale class systems, providing a missing link in general concept of collaborative workflow. \exacb{} enables CI-based benchmarking by
combining command-line tools, reusable CI/CD constructs such as GitLab CI/CD templates, and
established benchmarking harnesses -- currently using JUBE \cite{JUBE},  with planned support for ReFrame \cite{reframe} and Ramble \cite{jacobsen2023}. Several approaches to testing and benchmarking HPC applications have been proposed in the literature, including buildtest \cite{buildtest}, OLCF Test Harness \cite{OLCF}, Pavilion2 \cite{Pavilion2}, BenchPRO \cite{BenchPro}, Benchpark \cite{pearce2023}, and Thicket \cite{brink2023}, each tackling the aforementioned challenges in different ways. What distinguishes \exacb{} from these efforts is that its individual components are connected through a shared protocol and data model,
that allows standardized communication. This modular design enables components to
be exchanged as needed, supporting a wide range of applications and benchmarking scenarios in an extensible manner. Benchmark results are post-processed by the \exacb{} command-line tools to produce structured, machine-readable reports that can be easily analyzed and compared across systems and over time. The framework supports multiple benchmarking modes, including performance benchmarking, scalability studies, and energy efficiency evaluations. \exacb{} also allows easy extensibility by both developers and support engineers by creating and curating  a growing catalog of CI/CD components that can be used in existing pipelines.

A key design principle of \exacb{} is \emph{incremental adoption}. Benchmarks in a collection can
initially onboard at a minimal “runnability” level and then progressively evolve toward richer
capabilities such as instrumented measurements and full reproducibility. This approach lowers the
barrier to entry for application teams and enables rapid onboarding, while still providing an
auditable path toward producing structured results that can be tracked across system updates and
software changes.

We demonstrate the practicality of this approach through its application in JUREAP, an early-access
program preparing applications for the JUPITER exascale system. Using \exacb, we executed
benchmarks for a \emph{collection of over 70 applications} across diverse domains. Despite heterogeneous
maturity levels and an evolving system, runs were orchestrated automatically through shared CI/CD
components, and results were collected in a uniform format. This uniformity enabled advanced,
system-wide analyses with minimal additional effort. Beyond runtime metrics (time-to-solution),
\exacb{} supports framework-level workload augmentation, which can, for example, be used for non-invasive instrumentation by
injecting wrappers for energy measurement via shared platform configurations.

In parallel, we are integrating the \emph{JUPITER Benchmark Suite} \cite{herten24}, consisting of sixteen application
benchmarks and seven synthetic benchmarks with extensive JUBE workflows and reference results, into
\exacb. The JUPITER Benchmark Suite played a central role in the \emph{procurement of the JUPITER
system}, and its integration into \exacb{} will ensure that procurement-level benchmarks can be
reproduced continuously in CI/CD workflows. This ongoing effort brings the same declarative,
reproducible methodology to system-level evaluations of the JUPITER system,
thereby unifying application-centric studies with center-provided benchmark suites.

This work makes the following contributions:
\begin{enumerate}[leftmargin=*,itemsep=2pt,topsep=2pt]
  \item a declarative, CI/CD–centric framework (\exacb) that
    enables reproducible, large-scale benchmarking across heterogeneous collections  of
    benchmarks and systems;
  \item formaliztion of an incremental pathway for adoption (runnability
    $\rightarrow$ instrumentability $\rightarrow$ reproducibility) that
    lowers onboarding costs for benchmarks while preserving rigor;
  \item demonstration of automated execution and uniform data collection for a
    number of applications in JUREAP, enabling
    cross-application analyses and verifiable evaluation; and
  \item integratation of automatic experiments, like energy studies, via framework-level launcher injection and reporting on
    the ongoing onboarding of the JUPITER Benchmark Suite, which has
    already proven essential in the procurement process.
\end{enumerate}

\section{A Continuous Benchmarking Example with ExaCB}

To illustrate the concepts discussed in this paper, we present a practical example of continuous
benchmark using the \exacb{} framework with a simple application.

Before delving into the example, we first clarify the terminology used in this work. The terminology
in the field of benchmarking can be ambiguous, with terms like frameworks, environments and
harnesses often used interchangeably. To clarify our discussion, we define the following terms as
they are used in this work:

\begin{description}
  \item[Application] In HPC, an application is a scientific program to study a phenomenon, usually attached are a workload and fitting parameters.
  \item[Use Case] A use case is a specific use case of an application or a synthetic test. It typically includes input data, configuration parameters, and expected outputs. There can be multiple use cases for a benchmark, like the molecular simulation of a virus with 1 million or 8 million atoms.

  \item[Benchmark] A benchmark is a concrete, runnable implementation of a program, usually configured for simple, reproducible execution. Benchmarks may have multiple use cases, or also exactly one, in which case benchmark and use case may reduce solely to \emph{benchmark}.
    
  \item[Benchmark Harness] A benchmark harness is a tool that facilitates the execution of benchmarks under varying runtime conditions and parameter configurations. Example: JUBE.

  \item[Benchmarking Framework] A benchmarking framework, of which \exacb{} is an example, is an integrated collection of tools, rules, and protocols which allows one to manage running and execution of individual benchmarks and benchmark suites. Through alignment in a framework, its tools allow instrumentation of benchmarks as well as post-processing of benchmarks results.

  \item [Benchmark Suite] A benchmark suite is a thematically organized collection of benchmarks designed for systematic evaluation, often aligned to common rules. Typically, a benchmark suite includes tooling for analysis and comparison of results across its benchmarks to conduct a comprehensive multi-benchmark evaluation.
\end{description}

\subsection{Use Case: Logistic Map Application}

In this section, we consider a simple application called \texttt{logmap} \cite{logmap2011}, which computes the
logistic map function for a vector of input values. The application is designed to be a synthetic
benchmark with multiple use cases through varying the computational intensity and the workload. This allows simulating effects of different hardware parameters
like frequency scaling and number of cores used.

For the purposes of the example, it is enough to know the following characteristic of the
application:

\begin{itemize}
  \item The application can be compiled and installed using the command

\begin{boxshellnote}
cmake -S . -B build -DPROJECT_FEATURE=feature
cmake --build build
cmake --install build --prefix /opt/logmap/
\end{boxshellnote}

  \item The application can be executed using the command show below where \texttt{--workload-factor} and \texttt{--compute-intensity} are parameters that control the workload and computational intensity of the application respectively.

\begin{boxshellnote}
logmap --workload 6 --intensity 2.4
\end{boxshellnote}

\item The application outputs a file called \texttt{logmap.out} which contains the results of the computation along with the time taken to execute the application.

\item The application also outputs  a file called \texttt{logmap.stats} which contains the performance metrics of the application like execution time of the kernels.
\end{itemize}

\subsection{Benchmark: Logmap Benchmark in JUBE}

We now implement the benchmark for our use case using the JUBE benchmark harness.
JUBE is a Python-based tool which reads in a workload description script (\emph{JUBE script}) in XML or YAML and executes the defined commands. It resolves dependencies between individual commands and expands parameters, allowing for parameter space explorations through multiple definition of explored parameters. It can analyse execution output and comes with support for HPC job schedulers, like Slurm, through script inheritance functionality. By using tags, JUBE scripts can contain multiple execution and parameter definitions which are enabled or disabled during launch of the harness.
The logmap benchmark is
defined in a JUBE script called \texttt{logmap.yml}.
The JUBE script
initializes the parameters for the benchmark, sets up the environment, compiles the application, and
executes the application with different parameter configurations. The benchmark is written to
leverage JUBE's parameter study capabilities to run the application with different values of
\texttt{workload} and \texttt{intensity} as well as on different systems.

The benchmark can be executed with the JUBE harness per:

\begin{boxshellnote}
jube run logmap.yml --tags juwels-booster large-intensity large-workload
\end{boxshellnote}

The benchmark takes in two kinds of tags:
\begin{description}
  \item[System Name] The name of the system on which the benchmark is to be executed. In this case, we are executing the benchmark on JUWELS Booster. This enables setting system-specific options in the script.

  \item[Variant Tag] The variant tag specifies the benchmark variant to be executed. In this case, we are executing the \texttt{large-intensity} variant which corresponds to a \texttt{compute-intensity} of 2.4. The \texttt{large-workload} tag corresponds to a specific workload of the benchmark.
\end{description}

Through its analysis feature, JUBE produces a CSV file called \texttt{results.csv} after execution of all benchmarks which contains the columns described in \autoref{tab:result_columns}, per definition in the script. The information allows us to track the performance of the application across different systems and configurations, and over time.

\begin{table}[h]
\centering
\caption{Description of Result Columns}
\label{tab:result_columns}
\begin{tabular}{ll}
\toprule
\textbf{Column} & \textbf{Description} \\
\midrule
system           & system name \\
version          & system version \\
queue            & Slurm queue name \\
variant          & benchmark variant \\
jobid            & Slurm job ID \\
nodes            & number of nodes \\
taskspernode     & number of tasks per node \\
threadspertasks  & number of threads per task \\
runtime          & application-reported runtime \\
success          & correctness output \\
\textit{additional\_metrics} & \textit{user defined metrics} \\
\bottomrule
\end{tabular}
\end{table}

The required data is intentionally designed in such a way that system administrators can quickly identify jobs at a glance, while users get the performance metrics and execution status they need. The output structure is enforced in relatively rigid manner but can be easily relaxed to fit the need of the users. In this example, we focus on runtime as a performance metric, though many other metrics can be outputted. The result table shown here represents the minimum required output -- a baseline that stays consistent as users add more metrics via additional \textit{additional\_metrics} columns.

\subsection{ExaCB Framework Integration}

We now integrate the benchmark defined in the previous section into the \exacb{} framework to
enable continuous benchmarking on CI/CD platforms. In this case, we use \exacb's GitLab CI/CD
components to orchestrate the execution of the benchmark. In particular, we define a benchmark-specific GitLab CI/CD pipeline in a file called \texttt{.gitlab-ci.yml} which is shown here.

\begin{boxyamlnote}
include:
 - component: example/jube@v3.2
   inputs:
    prefix: "jedi.strong.tiny"
    variant: "large-intensity"
    machine: "jedi"
    queue: "all"
    project: "cjsc"
    budget: "zam"
    jube_file: "simple.yaml"
\end{boxyamlnote}

The GitLab CI/CD pipeline here calls the CI/CD component available as \texttt{example/jube@3.2}
which is a reusable component provided by \exacb{} to execute JUBE benchmarks on HPC systems. In this example, the
component uses the Jacamar \cite{ECP} runner to start a CI/CD job on the login node of the target HPC system
and sets up the directories and environment to execute the benchmark. During the setup of the
environment, the component also ensures that the compute account shown in the example is enabled so
that subsequent jobs can access the relevant partition.

The component then executes the benchmark using the JUBE command shown in the previous section. Once
the benchmark is executed, the component collects the results and uploads them to an orphan branch
called \texttt{exacb.data} in the benchmark repository. The results can then be visualized and
analyzed using \exacb's visualization tools or more featureful downstream tools.

\section{Design Principles and Related Work}

\begin{figure}
  \centering%
  \begin{tikzpicture}
    \tikzset{%
      quadrant box/.style={align=center, text width=7\cellwidth, align=center},
      quadrant label/.style={font=\itshape\small},
      number label/.style={shape=circle, fill=Gainsboro!50!white, text=white, font=\Large, inner sep=10pt}
    }
    \def\cellwidth{3}
    \def\cellheight{3}

    \draw[thick] (0,0) grid [step=\cellwidth] (2*\cellwidth, 2*\cellheight);

    \node[anchor=east, text=coupleyes, rotate=90, quadrant label] at (-0.3, 1.75*\cellheight) {Strong};
    \node[anchor=east, text=coupleno, rotate=90, quadrant label] at (-0.3, 0.75*\cellheight) {Loose};
    \node[anchor=north, rotate=90, text=couple, font=\bfseries, text width=16\cellheight, align=center] at (-1.5, \cellheight) {Degree of Benchmark+Harness Coupling};

    \node[anchor=north, text=centralyes, quadrant label] at (0.5*\cellwidth, -0.3) {Centralized};
    \node[anchor=north, text=centralno, quadrant label] at (1.5*\cellwidth, -0.3) {Uncentralized};
    \node[anchor=north, text=central, font=\bfseries, text width=20\cellwidth, align=center] at (\cellwidth, -0.8) {Degree of Collection Centralization};

    \coordinate (q1) at (0.5*\cellwidth, 1.5*\cellheight);
    \coordinate (q2) at (1.5*\cellwidth, 1.5*\cellheight);
    \coordinate (q3) at (0.5*\cellwidth, 0.5*\cellheight);
    \coordinate (q4) at (1.5*\cellwidth, 0.5*\cellheight);

    \node[quadrant box] at (q1) {\circled{1} \textcolor{centralyes}{Single repo}, \textcolor{coupleyes}{integrated harness}};
    \node[quadrant box] at (q2) {\circled{2} \textcolor{centralno}{Distributed repos}, \textcolor{coupleno}{external benchmarking harness with precise rules}};
    \node[quadrant box] at (q3) {\circled{3} \textcolor{centralyes}{Single repo}, \textcolor{coupleyes}{loose guidelines for benchmarks}};
    \node[quadrant box] at (q4) {\circled{4} \textcolor{centralno}{Distributed repos}, \textcolor{coupleno}{loose guidelines for benchmarks}};
  \end{tikzpicture}
  \caption{Categorization of benchmark collections and harness integration.}
  \label{fig:cat}
\end{figure}
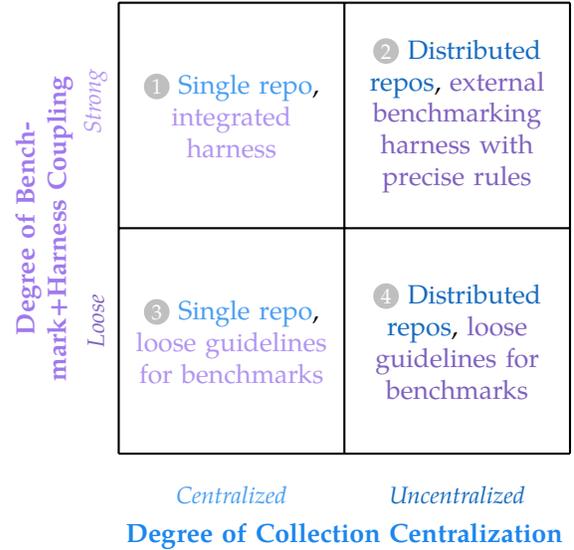

Collections of benchmarks may be integrated and connected to a shared harness in various fashions.
In order to understand the strengths of the design choices of \exacb, we categorize benchmark
collections along two dimensions: degree of \textcolor{central}{centralization of the collection of
benchmarks} and degree of \textcolor{couple}{coupling between benchmarks and harness}. The
combinations are summarized in \autoref{fig:cat} and explained in more detail in the following.

Benchmarks, collected in a central location, and tightly integrated to their executing harness
(\circled{1} in \autoref{fig:cat}), can leverage all benefits of directly-shared infrastructure and
move swiftly at a common pace. Direct embedding of the harness ensures immediate propagation of
enhancements and bugfixes. The common repository is a natural location for collection of harness and
benchmark artifacts, like benchmark results. The caveat of this approach is the created connection
between potentially disparate benchmarks including the associated user communities. Tight rules may
be introduced as a way of aligning benchmarks, which may be considered unattractive by creators of
individual programs. Extensibility of the single location is limited, as a contribution threshold is
introduced and the repository is guarded by a set of curators, which may create a bottleneck and
loss of control for benchmark creators. Non-main-repo extensions will always induce oversized forks,
limiting experimentation and ability to integrate into further use-cases of potentially
higher-level. In general, this approach sets the focus distinctly towards the collective nature of
the benchmarks, away from the developed programs under test.

If the single-location-collected benchmarks do not hold the benchmarking harness directly, but are
rather coupled more loosely to an external harness (\circled{3}), the alignment is rather on
common rules and guidelines. This allows the collection of benchmarks with arguably less effort,
but will also limit possibilities of (cross-) evaluation and impact. The harness may evolve
independently with higher pace and be used by a variety of individual collections, but no
immediate use-case is guaranteed, and the feedback limited. Extensions of the harness may need to
be incorporated (semi-)manually into the benchmarks, inducing delay or even omission. Benchmark
contributors may participate in contributing benchmarks more openly, though, as the impositions by
the harness may be less strict, and the contributors' typical execution patterns may need fewer
adaptions.

Building up on the loose guidelines of an external harness, distributing also the benchmarks across
multiple repositories relaxes the overall integration further (\circled{4}). The combination of
benchmarks and harness becomes rather a like-minded collective with loosely aligned targets. While
progress and synergistic effects are reduced, more freedom on both sides may improve progress and
identification with the endeavor. Common experiments, utilizing the diversity of the benchmarks, may
become more cumbersome to execute, reducing the potential impact.

Finally, distributing benchmarks along many collections, and enforcing a strong coupling to a
central, but external benchmark harness (\circled{2}), allows for individualized development of
benchmarks with distinct maintainers, with strong guidance and synergistic support by the harness.
Benchmarkers are enabled for independent experiments, but can easily participate in
collection-wide large-scale experiments through the harness. Enhancements and bugfixes can be
pushed forward quickly, allowing a swift development pace. As the harness enables well-aligned
artifacts, it may also add pre-defined artifact post-processing; by means of the enforced
protocol, artifacts may stay in the individual repositories for analysis by the contributor, or
move to a repository managed by the harness to evaluate the entirety of the collective benchmarks.
Benchmarks are still owned by individual contributors, but agreement on the strongly coupled
protocol creates a united benchmarking endeavor.

We consider the last-explained, strongly-coupled, but uncentralized approach of \circled{2} the most balanced to
achieve collective progress in a team of contributors, enabling code/performance reproducibility and
scientific discovery. We implement \exacb{} in this manner. Especially during onboarding, the coupling
of \exacb{} and benchmarks can also be relaxed, so that an ease of entry via~\circled{4} can be
offered, which is gradually moved towards~\circled{2}.

\section{\exacb{} Components}
\todo{Massage headings, text to fit "Components"}

The \exacb{} framework realizes a strongly coupled yet decentralized continuous benchmarking (CB)
infrastructure by building on established CI/CD technology and widely adopted software engineering
best practices. Rather than introducing a monolithic, centralized service, \exacb{} integrates
benchmark execution, data collection, and analysis into existing development workflows. At its core,
CI/CD orchestrators coordinate benchmark executions and post-processing steps, while users interact
with the framework through benchmark-specific Git repositories.

\subsection{Benchmark Repositories}
\label{exacb:overview:repo}

Each benchmark in \exacb{} is encapsulated in an individual Git repository, which serves as the
primary user-facing interface. While the framework is conceptually agnostic to the choice of version
control system, the current implementation focuses on Git due to its ubiquity and mature CI/CD ecosystem.

  Each benchmark repository includes at least one benchmark definition, implemented using a
  benchmarking harness that conforms to the \exacb{} protocol (see
  \autoref{exacb:overview:protocol}). For maximal reproducible, the benchmark definition is a
  source-based implementation that includes all necessary build and execution instructions.

  However, in cases where maximal reproducibility is not yet feasible or desired, it is also possible for the
  benchmark to directly reference pre-built binaries or external resources.

Central to each repository is a CI/CD configuration file that defines a \emph{benchmark pipeline}.
While \exacb{} is designed to be CI/CD-agnostic, the current implementation supports GitLab CI
through the \texttt{.gitlab-ci.yml} configuration format. Support for additional CI/CD systems, such
as GitHub Actions, is planned as future work. The CI configuration connects repository-specific
execution logic to the \exacb{} back-end via the protocol, enabling standardized orchestration and
data exchange.

To execute the benchmarks on a target system,  the repository must be associated with at least one
CI runner capable of accessing the tested HPC environment. There are multiple different runners
that can be used in this regard, with Jacamar and Hubcast~\cite{Hubcast} being some of
the popular ones. \exacb{} currently supports Jacamar with plans to support additional runners in
the future.

Typically, benchmark authors (who can either be application developers or support staff at the
supercomputing centres) are responsible for creating and maintaining benchmark repositories. To
support this process, \exacb{} provides documentation, examples, and best-practice guidelines, as well as interactive training sessions.
Benchmark repositories may be organized into collection-specific groups, forming a semi-centralized
structure, or remain entirely independent. In any case, the \exacb{} implementation is external to these benchmark repositories.

\subsection{\exacb{} Protocol}
\label{exacb:overview:protocol}

To enable strong coupling between the independently developed benchmarks and the \exacb{} runtime, the
framework defines a standardized interfacing protocol. This protocol connects the benchmark
repositories (\emph{front end}) with the \exacb{} orchestration and analysis components
(\emph{back end}) in a well-defined manner.

The protocol specifies experiment structures, configuration parameters, data formats, metadata
schemas, and naming conventions. It spans multiple layers of the system: CI/CD configuration files
trigger experiment workflows, the \exacb{} back end invokes repository-specific benchmark execution
logic, and post-processing components collect, validate, and store benchmark results. By enforcing
these conventions, the protocol enables interoperability and reproducibility across heterogeneous
benchmark collections.

\subsection{\exacb{} Structure}
\label{exacb:overview:implementation}
\todo{Another title for this (because there's a section "implementation" already!}

The \exacb{} implementation centers on the orchestration of benchmark executions as defined by the
protocol. Building on reusable CI/CD components, currently GitLab CI/CD components, the framework
leverages CI runners deployed on HPC systems to launch benchmarks, retrieve their outputs, perform
post-processing, and propagate results to storage.

Internally, \exacb{} employs a combination of standard tools and lightweight custom components,
including Git command-line interactions and a small auxiliary Python tool. Each stage of the
workflow is realized as an individual CI job. The jobs communicate between themselves primarily
through the CI/CD's native artifact management capabilities, and optionally through external storage
back ends.

This design allows \exacb{} to exploit existing CI/CD infrastructure, traditionally used for
software testing and integration, for continuous benchmarking purposes. The reuse of established
CI/CD technology enables application developers and system administrators to leverage
familiar tools and practices, lowering the barriers for collaboration.

The modular structure further enables coordinated execution of benchmarks across multiple
repositories through cross-triggered CI pipelines.

\subsection{Benchmark Execution}
\label{exacb:overview:execution}

Benchmark execution is initiated when a CI pipeline is triggered, either manually, on a schedule, or
by an eligible repository event. The \exacb{} back end interprets experiment definitions provided in
the repository’s CI/CD configuration and dispatches corresponding jobs to the CI runner, which in
turn submits workload jobs to the HPC system’s batch scheduler.

\exacb{} itself does not directly submit or evaluate HPC jobs. Instead, it orchestrates the
continuous benchmarking workflow and delegates execution to an external benchmarking harness that
conforms to the protocol and provides an appropriate adapter. The framework is designed with
established harnesses in mind, such as JUBE, ReFrame, or Ramble. In the benchmark collections presented in this work, JUBE is used exclusively.

Through its JUBE adapter, \exacb{} instantiates run definitions from template JUBE scripts and
relies on JUBE’s Slurm integration to execute benchmarks or parameter studies on the HPC system.
JUBE monitors job completion and performs initial analysis of the output, which is then ingested by
\exacb{} for further post-processing.

The inclusion of a benchmarking harness as an explicit protocol component enables reuse of
established tooling for one-off benchmarks while making results consumable within the continuous
benchmarking workflow. This integration also facilitates back-end-driven analysis and experiment
extensions , and allows users to leverage benchmarking tools beyond
CI-driven scenarios.

\subsection{Data Collection}
\label{exacb:overview:data}

Data collection in \exacb{} primarily relies on the structured output produced by the benchmarking
harness. Since the harness is part of the protocol, benchmark results are well-formed and
machine-readable. Depending on the benchmark collection and experiment scope, the set of recorded
metrics may vary; commonly reported values include application runtimes.

The \exacb{} framework parses the harness output, applies sanity checks and post-processing steps,
and feeds the resulting data back into the CI pipeline. Collected data are associated with the
corresponding CI jobs as artifacts and may additionally be stored in persistent locations, such as
orphaned Git branches or dedicated object storage (e.g., S3-based back ends).

For analysis purposes, \exacb{} can incorporate externally provided data through well-defined
injection hooks. While this allows augmentation with manually curated or third-party data, the
resulting chain of trust is not guaranteed.

\subsection{Data Analysis}
\label{exacb:overview:analysis}

To decouple execution and data acquisition from evaluation, \exacb{} provides dedicated CI jobs for
data analysis. Analysis targets may include individual benchmark runs, aggregated benchmark
collections, or combinations thereof, including injected external data. Due to the standardized data
formats enforced by the protocol, the analysis pipeline can also be applied outside of a full
\exacb{} workflow. Analysis may be arbitrarily complex, therefore \exacb{} ensures a proper data storage format, and itself only provides lightweight analysis.

The analysis in \exacb{} is facilitated through a dedicated Python package with a command-line interface,
developed primarily for the experiments presented in this work. The analysis tooling can be extended
or replaced to accommodate the needs of different benchmark collections. Evaluated data are stored
analogously to raw results, as CI artifacts, in versioned branches, or in external storage, and may
be emitted in machine-readable formats or as human-readable reports.

By retaining versioned benchmark outputs, \exacb{} enables comprehensive and even a-posteriori
time-series analyses, supporting long-term performance tracking, trend identification, and
regression detection. Aggregated results can further be exported to external monitoring and
visualization systems, such as Grafana~\cite{Grafana} or LLview~\cite{llview}, facilitating
integrated analyses across multiple data sources.

\section{Implementation}

\exacb{} is implemented in terms of a collection of tools and protocols to support
integration at various levels of the code. The development of the tools and protocols in \exacb{} is
an ongoing effort, with new features and capabilities being added over time. In this section, we
illustrate the current state of the implementation of \exacb{} by focusing on some core components of
the framework.

\subsection{CI/CD Orchestrators}
\label{sec:orchestrators}

The core of the framework are the \exacb{} CI/CD orchestrators, which are used to manage the execution of benchmarks and then process and store the results. In the current state of the implementation, the CI/CD orchestrators are currently implemented using GitLab CI/CD components.

Large benchmarks often demand substantial compute resources, yet access to these resources might be scarce and depend on auxiliary infrastructure such
as cloud storage for inputs, caches and results. These supporting services may experience transient failures during execution.

To prevent costly re-execution and ensure result recovery, \exacb{} avoids a monolithic orchestrator and instead uses multiple independent orchestrators to manage execution and post-processing. This
separation allows benchmark progress and results to be preserved despite partial infrastructure
failures. In current state, there are three main orchestrators: the Execution, the Feature-Injeciton and the Post-Processing Orchestrator.

\subsubsection{The Execution Orchestrator}

The execution orchestrator is responsible for managing the execution of benchmarks on various
systems. It sets up the environment for the execution, triggers the execution, and collects the results of the benchmark. It then forwards the results to the post-processing
orchestrator for further processing. For GitLab, the execution orchestrator is implemented as a
GitLab CI/CD template, with exploration going on for using CI/CD Steps and Github Actions in future.

\begin{boxyamlnote}
- component: execution@v3
  inputs:
    prefix: "jureca.single"
    # Benchmark specification
    usecase: "bigproblem"
    variant: "single"
    jube_file: "benchmark/jube/shell.yml"
    # Machine and queue specification
    machine: "jureca"
    queue: "dc-gpu"
    project: "cexalab"
    budget: "exalab"
    fixture: .setup
    record: "true"
\end{boxyamlnote}

In the above example, the user implements the component in their GitLab CI/CD pipeline
\texttt{.gitlab-ci.yml} file to invoke the execution orchestrator. Here, the execution orchestrator
is a JUBE-based orchestrator that executes benchmarks using the JUBE benchmarking harness. The
orchestrator looks for the benchmark definition in the \texttt{jube\_file} input key. The
orchestrator expects the benchmark to be implemented as a JUBE benchmark and supporting
customization of the benchmark via tags such as the \texttt{variant} and \texttt{usecase}  keys. The \texttt{variant} key is an example of a strong degree of coupling between the application and the harness. The value of this key can be strictly enforced and could, for instance, correspond to the scale or size of a solved problem that is shared with the same values by different applications to be benchmarked. On the other hand, \texttt{usecase} defines application specific tags, reflected in underlying definition of a JUBE script.

The GitLab invoked with a set of inputs that define the execution environment and the benchmark to
be executed. In this case, the execution orchestrator specifies that all the CI jobs that are
created should be prefixed with \texttt{jureca.single}. The benchmark is executed on the
\texttt{jureca} machine in the \texttt{dc-gpu} queue using the \texttt{cexalab} project and the
\texttt{exalab} budget.

Often, benchmarks require setup and teardown steps to prepare the environment and assets for
execution, and then clean up after execution. The \texttt{fixture} input allows the user to specify a
CI/CD job template that is executed before and after the benchmark execution. Finally, the
\texttt{record} input specifies whether the results of the benchmark execution should
be recorded. When set to \texttt{true}, the results are stored in either a cloud storage or an
orphan branch in the GitLab repository of the benchmark named \texttt{exacb.data}.

\subsubsection{The Post-Processing Orchestrator}

The post-processing orchestrator is responsible for analyzing and processing the results collected
by the execution orchestrator. It takes the raw results and transforms them into a structured,
machine-readable format that can be easily analyzed and compared across systems and over time.
Depending on the requirements, different post-processing orchestrators can be implemented to support
various analysis and processing tasks. In the current state of the implementation, the
\exacb{} framework supports three main post-processing orchestrators:

\begin{itemize}
  \item Machine comparison

      The machine comparison orchestrator compares the performance of the benchmark across
      different systems and configurations and plots the results.

  \item Scalability analysis

      The scalability analysis orchestrator analyzes the scalability of the benchmark across a
      single system and plots the results.

  \item Time-series analysis

      The time-series analysis orchestrator plots the values of given preformance metric over time.
      
\end{itemize}

In this section, we describe the machine comparison post-processing orchestrator as an example.
As with the execution orchestrator, the post-processing orchestrator is implemented as a GitLab
CI/CD template.

\begin{boxyamlnote}
- component: comparison@v3
  inputs:
    prefix: "evaluation.jedi"
    pipeline: [ "221622", "221622" ]
    selector: [ "jedi.evaluation.jedi" , "jedi.evaluation.jureca" ]
\end{boxyamlnote}

In the above example, the user implements the text above in their GitLab CI/CD pipeline
\texttt{.gitlab-ci.yml} file to invoke the machine comparison post-processing orchestrator. Here, the orchestrator compares the results of benchmarks executed in two different pipelines
\texttt{221622} on two different systems \texttt{jedi} and
\texttt{jureca} respectively. The \texttt{selector} input specifies the prefixes of the CI jobs
whose results are to be compared. The \texttt{prefix} input specifies that all the CI jobs that are
created should be prefixed with \texttt{evaluation.jedi}. The post-processing orchestrator retrieves
the results from the \texttt{exacb.data} branch of the benchmark repositories and processes them to
generate comparison plots ...

\begin{boxyamlnote}
- component: time-series@v3
  inputs:
    prefix: "jupiter.benchmark.stream.cuda"
    pipeline: [] # Optional -- empty take "all"
    data_labels: [ "Copy BW  [MBytes/sec]", "Mul BW [MBytes/sec]", "Add BW [MBytes/sec]", "Triad BW [MBytes/sec]", "Dot BW [MBytes/sec]" ]
    ylabel: [ "Bandwidth / MB/s"]
    plot_labels: [ "Copy kernel", "Multiply kernel", "Add kernel", "Triad kernel", "Dot kernel" ] # Optional -- empty take data_labels as plot_labels
    time_span: [ "2026-01-01", "2026-04-01" ] # Optional -- empty take "all"
    ...
\end{boxyamlnote}

The time-series plotting component of \exacb{} converts generated output into continuous visualizations of selected performance metrics. This enables users to assess system health and the impact of software stack updates, upstream application changes, or hardware modifications. For this component, shown above, \texttt{prefix} of the considered experiment must be defined and \texttt{pipeline} input can be defined to specify the selected pipelines. \texttt{pipeline} input is an optional component, meaning that if it is not defined, the component will choose all the pipelines corresponding to given \texttt{prefix}. The performance metrics to be plotted are defined using the \texttt{data\_labels} input, which selects the columns of the resulting data with the names provided for this input. To customize the resulting plot, the user defines the \texttt{ylabel} and optionally \texttt{plot\_labels}; if it is not defined, \texttt{data\_labels} will be used as \texttt{plot\_labels}. To plot data from a given time period, the user can define the optional input \texttt{time\_span}; again f not defined, all available data will be plotted.

\subsubsection{The Feature-Injection Orchestrator} The feature-injection orchestrator allows users to run additional test cases on applications directly directly based on existing, fixed definition JUBE file for single tests by injecting a command.

\begin{boxyamlnote}
- component: feature-injeciton@v3
  inputs:
    prefix: "jupiter.single"
    usecase: "problem"
    variant: "single"
    jube_file: "benchmark/jube/shell.yml"
    ...
    in_command: "export UCX_RNDV_THRESH=intra:65536,inter:65536"
\end{boxyamlnote}

Using this orchestrator it is also possible to inject a feature into the existing benchmark/analysis. The advantage of such approach is the re-usability of predefined benchmark, here using JUBE framework, and creating a full analysis independent of JUBE definition. This provides a consistency between continuously running benchmarks while still allowing to run special tests -- in the example above changing values of \texttt{UCX\_RNDV\_THRESH} for single test.

The separation of execution and post-processing orchestrators allows for greater flexibility and resilience in the benchmarking process. The execution orchestrator can focus on managing the execution of benchmarks, while the post-processing orchestrator can focus on analyzing and processing the results. It also allows development of multiple post-processing orchestrators, used depending on the requirements of the benchmarking process, without having to rerun the benchmarks themselves. Introduced components encapsulate many different types of workloads, allowing to run from basic benchmarks, data analysis and plotting to precise investigation into parameters.

\subsection{The \exacb{} Protocol}

As described in \autoref{sec:orchestrators}, the \exacb{} framework consists of multiple
independent components that either generate or consume benchmark data at various stages of the
benchmarking process. In particular, the generation and the consumption of benchmark data are
completely decoupled and may occur at different points in time or on different systems. To enable
this decoupling, the framework relies on a shared protocol and data model that allows all components
to communicate in a standardized, machine-readable manner.

The \exacb{} protocol is designed to capture the output of benchmark executions in a structured,
self-describing format. It follows a hierarchical data model expressed in JSON and is intended to
be extensible, reproducible, and robust against partial or incremental data generation. Each
protocol document represents a single benchmark report and consists of the following top-level
sections:

\paragraph{Version}
A protocol version identifier that specifies the schema version used to encode the benchmark
report. This enables backward compatibility and schema evolution.

\paragraph{Reporter}
Metadata describing the entity that generated the report. This includes information about the
generator tool, the execution environment, and the provenance of the data, such as pipeline and job
identifiers, commit hashes, user information, system name, software version, and timestamps. This
section ensures traceability and reproducibility of benchmark results.

\paragraph{Parameter}
A (possibly empty) collection of global benchmark parameters that apply to the entire report. These
parameters are intended for experiment-wide configuration values that are not specific to a single
execution instance.

\paragraph{Experiment}
A description of the experimental context under which the benchmark was executed. This includes
the target system, software version, experiment variant, and a timestamp indicating when the
experiment was initiated. This section provides semantic context for interpreting the benchmark
results.

\paragraph{Data}
An array of benchmark result entries, where each entry corresponds to a single benchmark execution
or run. Each data entry captures both configuration and outcome information, including:
\begin{itemize}
    \item execution success status,
    \item total runtime,
    \item execution parameters such as number of nodes, tasks per node, and threads per task,
    \item scheduler-specific metadata (e.g., job identifiers and queue names),
    \item an extensible metrics object for benchmark-specific performance data.
\end{itemize}

The separation between metadata, experimental context, and execution data allows benchmark results
to be produced, stored, and analyzed independently of the components that generated them. By using
a flexible, self-describing data format, the \exacb{} protocol supports heterogeneous benchmark
workloads, diverse execution environments, and future extensions without requiring changes to
existing consumers of the data.

Overall, the \exacb{} protocol provides a unified and extensible foundation for representing
benchmark outputs in a reproducible and interoperable manner across the entire \exacb{} framework.

\subsection{Auxiliary Tools}

In addition to the CI/CD orchestrators and the core \exacb{} protocol, the framework is supported by
a collection of auxiliary tools intended to facilitate common benchmarking-related tasks. These
tools are primarily implemented in Python and are designed to operate independently of the
orchestration layer, consuming protocol-compliant benchmark data as input.

At the current stage of development, only a subset of the auxiliary tools is fully implemented and
actively used within the \exacb{} workflow. The implemented auxiliary tools focus mainly on post-hoc analysis and visualization of benchmark results. They provide functionality for parsing protocol documents, aggregating results across multiple executions, and generating plots for performance comparison and scalability analysis.

Additional auxiliary tools are currently work in progress. These include utilities for deeper
integration with CI/CD systems, automated retrieval of benchmark reports from artifact and storage
back ends, and extended validation of protocol compliance. While not yet part of the stable tool set,
their interfaces and design are guided by the same protocol-driven principles as the rest of the
framework.

Overall, the auxiliary tools are intended to complement the orchestrators and the protocol by
enabling flexible and reproducible analysis workflows. As the \exacb{} framework matures, these
tools are expected to evolve from experimental or planned components into stable, reusable building
blocks, without requiring changes to the core protocol or orchestration logic.

\section{Experiments and Results}

In this section, we present experimental results obtained using \exacb{} in the
context of different experiments on the JUPITER supercomputer at Jülich Supercomputing
Centre (JSC). These experiments demonstrate the capabilities of \exacb{} in orchestrating large-scale
benchmarking collections, integrating energy measurements, enabling cross-application analyses, and
supporting reproducibility evaluation.

\subsection{Performance Measurements}

A central goal of JUREAP was to continuously evaluate the JUPITER machine and the early-access
applications on it at a large scale, rather than limiting experiments to a small set of reference
workloads. In our current deployment, we were able to run continuous benchmarks for a
\emph{collection of almost thirty applications} from the JUREAP portfolio. These cover diverse
scientific domains and represent a variety of workflows, from lightweight test cases to
production-scale simulations.

Importantly, these benchmarks were executed automatically through \exacb{}. Some benchmarks in the collection were only at the runnability stage, others already provided instrumentation, and a subset had reached full reproducibility. The ability to incorporate various benchmarks at diverse level of preparedness was crucial to carry out evaluation of the system and the applications in a short amount of time. 

This demonstrates the value of the \textbf{protocol + implementation}
design: the protocol ensured that results from all benchmarks—regardless of maturity—were captured
in the same standardized format, while the implementation provided the shared components for
orchestration and collection. This allowed the collection to be tracked as a whole, enabling
system-wide analysis even when benchmarks were heterogeneous in maturity.

\begin{figure}
  \centering
  \includegraphics[width=\linewidth]{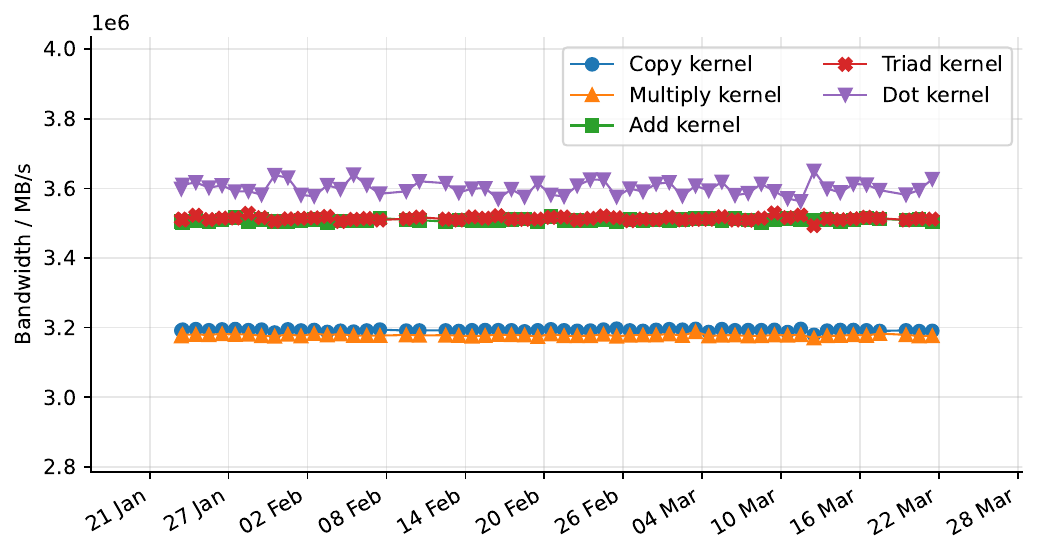}
  \caption{Post-processing orchestrator plotting results of BabelStream (GPU) over time.}
  \label{fig:stream-timeseries}
\end{figure}

\begin{figure}
  \centering
  \includegraphics[width=\linewidth]{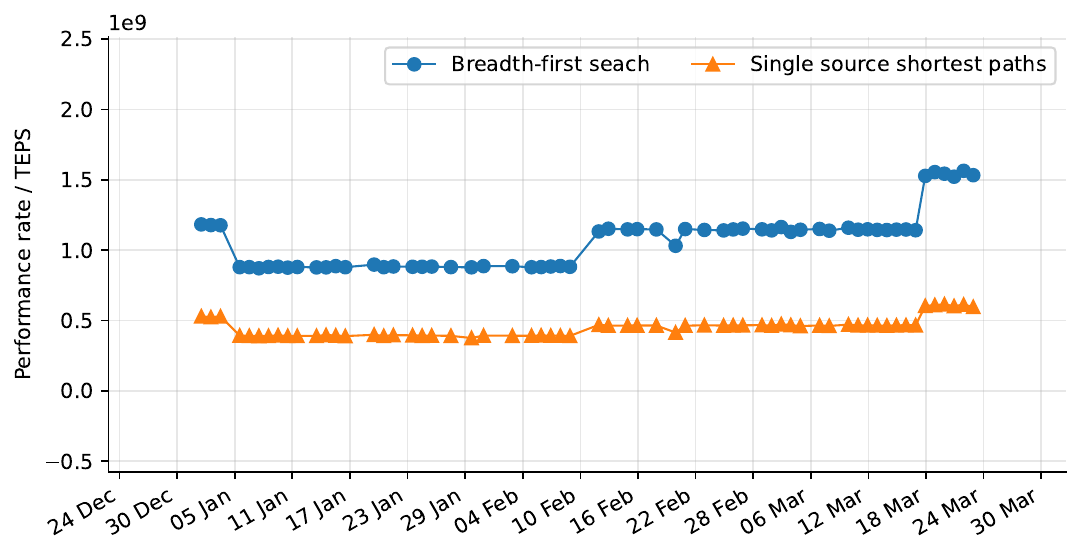}
  \caption{Post-processing orchestrator plotting results of GRAPH500 over time.}
  \label{fig:graph500_timeseries}
\end{figure}

One of the examples utilizing the introduced framework and time-series component in order to monitor system behavior is shown in \autoref{fig:stream-timeseries}. The BabelStream benchmark~\cite{babelstream2018} to collect memory bandwidths is executed on daily scheduled pipeline and the resulting bandwidth is plotted against time. Another time-series is shown in \autoref{fig:graph500_timeseries}, which displays the performance of two kernels of the GRAPH500 benchmark~\cite{ang2010graph500}, again executed on daily basis. Comparing the two time-series plots, it can be seen that performance of BabelStream remains constant while GRAPH500 has visible changes to its performance due to system changes. These examples allows users, support staff, and administrators to immediately understand performance of the system components (memory sub-system, inter-node communication, I/O and network fabric) and spot either performance stability (\autoref{fig:stream-timeseries}) or regression/recovery (\autoref{fig:graph500_timeseries}).

\begin{figure}
  \centering
  \includegraphics[width=\linewidth]{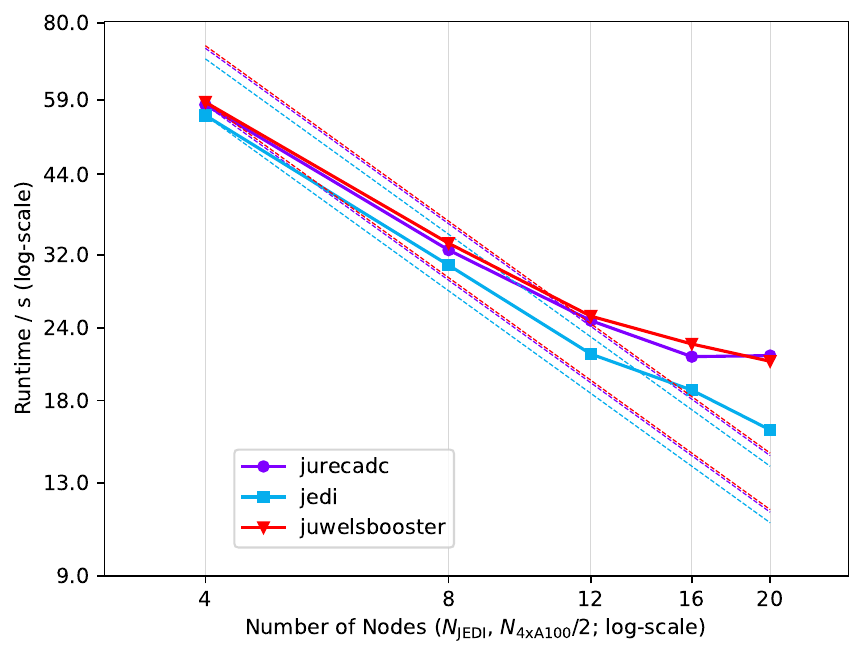}
  \caption{Post-processing orchestrator plotting results showing comparison of benchmark results between JEDI, JUWELS Booster and JURECA-DC.}
  \label{fig:comparison-plot}
\end{figure}

Another example of performance measurement is shown in \autoref{fig:comparison-plot} where the runtime of an application (time to solution) is  plotted against the number of nodes on different systems available at JSC, utilizing the machine component of \exacb{}. The shown strong-scaling comparison is across two different GPU generations (NVIDIA Ampere and NVIDIA Hopper), and high-performance network types; the Ampere result is halved for easier comparability). Superimposed are scaling bands to guide the eye in the \qty{80}{\percent} scaling regime. The plot allows users to immediately understand scaling performance, including GPU-generational and system-specific differences. The definition for this post-processing orchestrator can be generated once, and then rolled out to all benchmarks participating in this study, for rapid assessment.

\begin{figure}
  \centering
  \includegraphics[width=\linewidth]{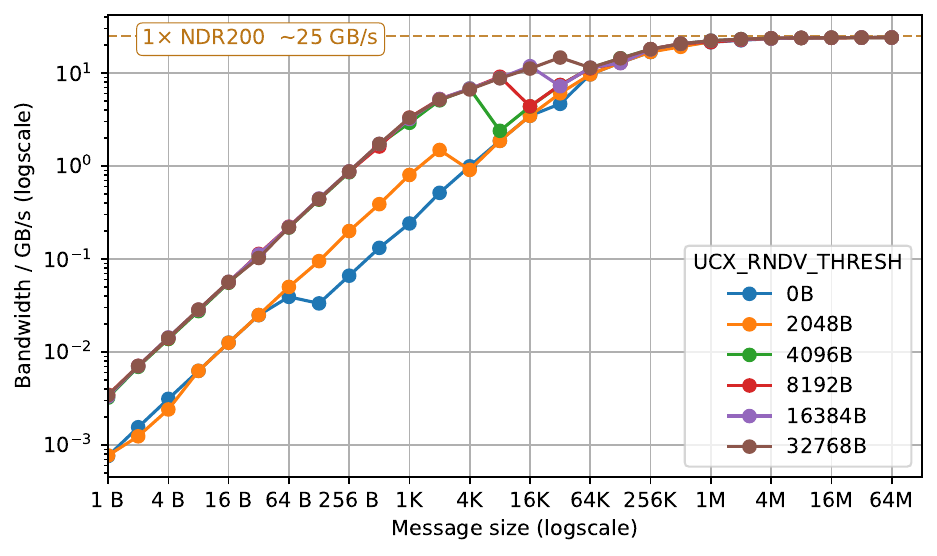}
  \caption{Post-processing orchestrator plotting bandwidth results of OSU Microbenchmark against message size.}
  \label{fig:osu_bw}
\end{figure}

In \autoref{fig:osu_bw}, the feature injection support of the presented framework is used. In this case, the OSU Microbenchmarks \cite{osu_bench}, testing inter-node MPI performance, are evaluated under different configuration values of the underlying UCX framework \cite{shamis2015ucx}, \texttt{UCX\_RNDV\_THRESH}. The different values are injected to the \exacb-baesd pipeline without changing the actual benchmark. The plot shows the resulting bandwidth against the message size for six example values. Through evaluations like this, both global system setting defaults for general workloads, and individual, application-specific values can be analyzed. Through this, finding the ideal working spot for individual applications is only one pre-defined experiment away.

\begin{figure}
    \centering
    \includegraphics[width=\linewidth]{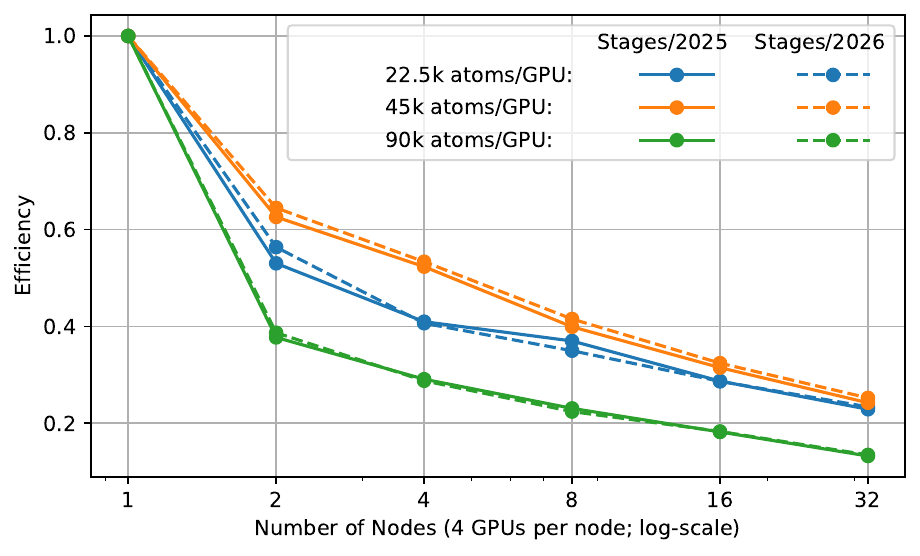}
    \caption{Weak scaling of a benchmark application.}
    \label{fig:app_weak_scaling}
\end{figure}

Two further pre-defined experiments enable analysis of software updates in the supercomputer environment and evaluation of weak-scaling efficiency. In \autoref{fig:app_weak_scaling}, we combine both experiments in one plot. The application is run across a number of nodes with size-adapted workloads, for two different sets of software dependency, here aligned into software \emph{stages} {2025 vs. 2026}. The individual components shown allow users to guide updates of the application stack, ensuring no performance regressions occur or can be solved immediately, and understand weak-scaling capacity of their application to match expectations, as weak-scaling behavior is an inherent application-specific feature. Together, the combined plot exemplifies that also complex, multi-faceted data can be presented effectively to users, giving agency about their individual workloads.

\subsection{Energy Studies}

Energy efficiency is a key performance dimension for exascale systems.
In \exacb{}, energy measurements are obtained by running benchmarks through
the energy-aware launcher \texttt{jpwr}. This support is typically enabled
without modifying the benchmarks themselves, similar to \autoref{fig:osu_bw}. The JUBE platform configuration
selects \texttt{jpwr} as the launcher, and the \texttt{jureap/energy} component
in the CI/CD pipeline is activated to collect and export the corresponding
energy-to-solution data.

Within \exacb{}, this maps to the \emph{implementation} layer. The protocol
defines how energy-to-solution must be represented in the results, while the
implementation provides the concrete reporting templates and launcher-based data
collection needed to produce it. Crucially, this additional functionality could
be introduced without changing the benchmarking code. The benchmarks remain
unchanged, yet the resulting dataset is enriched with protocol-compliant energy
information.

\begin{figure}[h]
  \centering
  \includegraphics[width=\linewidth]{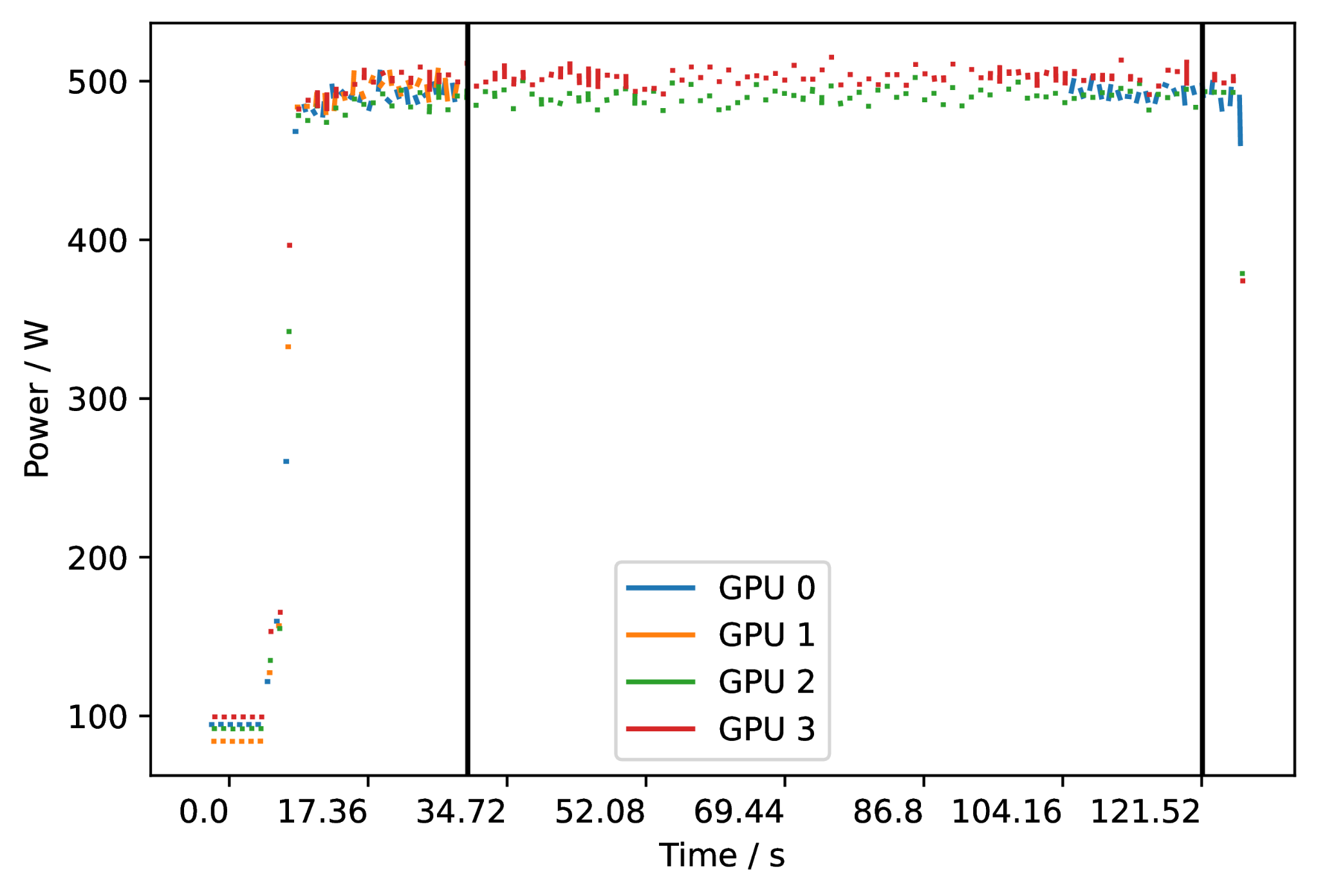}
  \caption{Energy-to-solution measurement visualization for a benchmark in JUREAP.}
  \label{fig:energy}
\end{figure}    

\begin{figure}[h]
  \centering
    \includegraphics[width=0.44\textwidth]{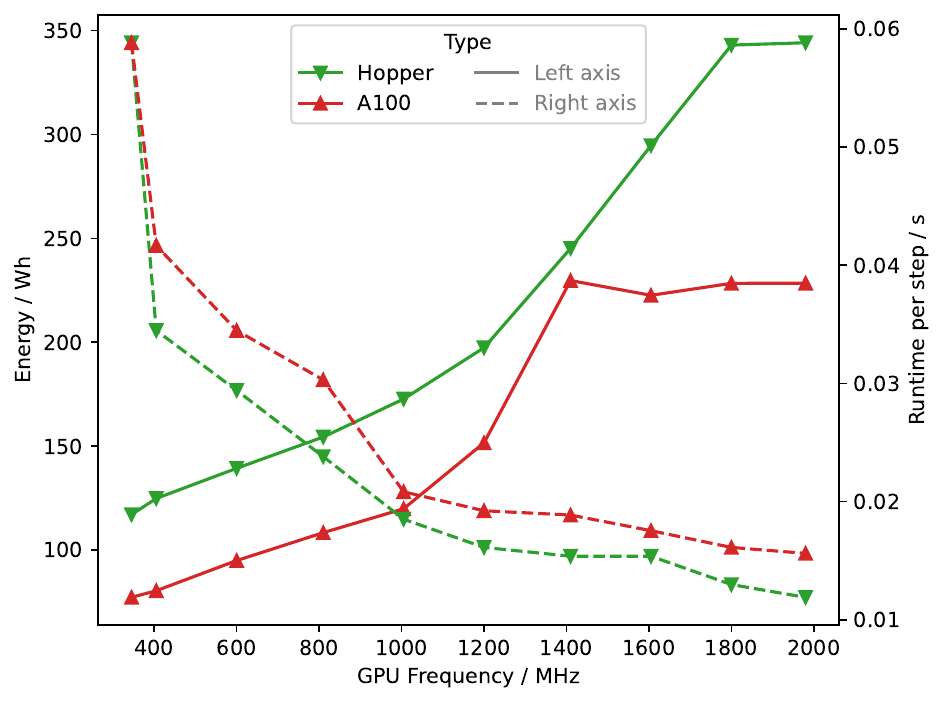} \includegraphics[width=0.44\textwidth]{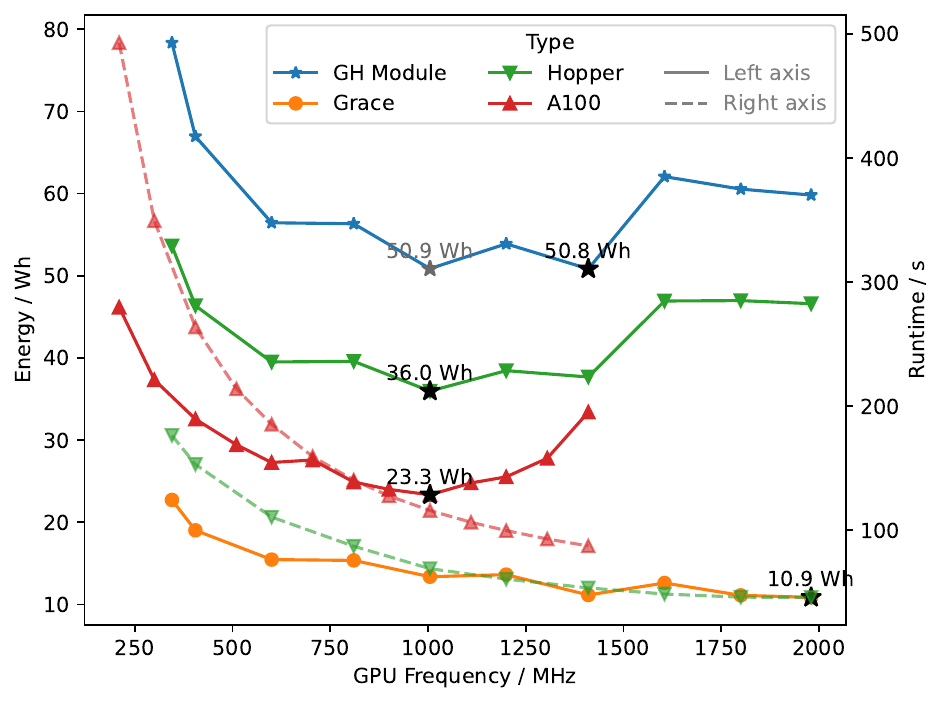}
  \caption{\exacb-enabled energy measurements of two JUREAP applications for a variety of CPU/GPU operation frequencies, determining sweet-spots for energy-efficient operation}
  \label{fig:jedi-energy}
\end{figure}

\autoref{fig:energy} illustrates energy-to-solution results for benchmarks for an example application from JUREAP. The power values of the foru GPUs are shown, together with black vertical bars which indicate the scope of the measurement. The measurement scope excludes start-up and wind-down phases, as they are in many cases not representative of the overall application profile -- of course, this systematically underestimates the reported energy. The semi-automatic approach automatically places the vertical guide, but allows for human verification and adaption. In \autoref{fig:jedi-energy}, results of such calibrated energy measurements are shown under variation of the processor frequency, finding sweet spots for energy-efficient operation. The approach is scalable also hundreds of jobs.

These results show that energy-aware benchmarking in JUREAP can be integrated
seamlessly into continuous workflows, requiring minimal effort from application
developers while providing deep insight into performance and efficiency.

\section{Conclusion}

We presented \exacb{}, a continuous benchmarking framework that
integrates performance evaluation into CI/CD workflows for HPC systems,
making benchmarking a first-class component of the software development
process. The framework establishes a flexible and non-intrusive pipeline
spanning execution, instrumentation, analysis, and data management,
supported by reusable components, standardised harnesses, and shared
reporting protocols.

\exacb{} supports incremental adoption, allowing applications to onboard
with minimal effort and progressively advance toward more sophisticated
instrumentation and reproducibility. This enables consistent benchmarking
across applications at different levels of maturity while maintaining a
unified methodology and clear separation of concerns.

The framework has been operationally deployed in the JUPITER Research
and Early Access Program (JUREAP), where it has supported continuous
benchmarking of more than 70 HPC applications. In practice, this enables
performance monitoring, long-term regression tracking, and systematic
parameter exploration as part of everyday development and operation.

These results demonstrate that continuous benchmarking is practical even
for large and heterogeneous application collections. Beyond site-level
deployment, the approach provides a foundation for extending benchmarking
to broader research workloads and collaborative environments.

Overall, \exacb{} elevates benchmarking from an ad hoc activity to a
continuous, structured software engineering practice by making it an
integral part of CI/CD-driven development workflows.

Future work will extend the current framework by strengthening its
software engineering foundations, including formalising workflows as
composable pipelines, integrating performance and energy evaluation into
testing and verification with support for long-term regression analysis,
and improving reproducibility through enhanced configuration management
and provenance tracking. We will further expand CI integration beyond
the current setup to include GitHub Actions and additional CI/CD
frameworks, enabling wider adoption without compromising the continuous
benchmarking model.

\section*{Acknowledgment}
\addcontentsline{toc}{section}{Acknowledgment}

\printbibliography

@inproceedings{jacobsen2023,
  title={Ramble: A flexible, extensible, and composable experimentation framework},
  author={Jacobsen, Douglas and Bird, Robert F},
  booktitle={Proceedings of the SC'23 Workshops of The International Conference on High Performance Computing, Network, Storage, and Analysis},
  pages={600--608},
  year={2023}
}

@inproceedings{koskela2023,
  title={Principles for automated and reproducible benchmarking},
  author={Koskela, Tuomas and Christidi, Ilektra and Giordano, Mos{\`e} and Dubrovska, Emily and Quinn, Jamie and Maynard, Christopher and Case, Dave and Olgu, Kaan and Deakin, Tom},
  booktitle={Proceedings of the SC'23 Workshops of The International Conference on High Performance Computing, Network, Storage, and Analysis},
  pages={609--618},
  year={2023}
}

@misc{ECP,
  author = {ECP},
  title = {{Jacamar CI}},
  howpublished = "\url{https://gitlab.com/ecp-ci/jacamar-ci}",
  year = {2010}, 
  note = "[Online; accessed 19-July-2008]"
}

@inproceedings{pearce2023,
  title={Towards collaborative continuous benchmarking for HPC},
  author={Pearce, Olga and Scott, Alec and Becker, Gregory and Haque, Riyaz and Hanford, Nathan and Brink, Stephanie and Jacobsen, Doug and Poxon, Heidi and Domke, Jens and Gamblin, Todd},
  booktitle={Proceedings of the SC'23 Workshops of The International Conference on High Performance Computing, Network, Storage, and Analysis},
  pages={627--635},
  year={2023}
}

@inproceedings{brink2023,
author = {Brink, Stephanie and McKinsey, Michael and Boehme, David and Scully-Allison, Connor and Lumsden, Ian and Hawkins, Daryl and Burgess, Treece and Lama, Vanessa and L\"{u}ttgau, Jakob and Isaacs, Katherine E. and Taufer, Michela and Pearce, Olga},
title = {Thicket: Seeing the Performance Experiment Forest for the Individual Run Trees},
year = {2023},
isbn = {9798400701559},
publisher = {Association for Computing Machinery},
address = {New York, NY, USA},
url = {https://doi.org/10.1145/3588195.3592989},
doi = {10.1145/3588195.3592989},
booktitle = {Proceedings of the 32nd International Symposium on High-Performance Parallel and Distributed Computing},
pages = {281–293},
numpages = {13},
location = {Orlando, FL, USA},
series = {HPDC '23}
}

@article{allen2020,
  doi = {10.5281/ZENODO.4324415},
  
  url = {https://zenodo.org/record/4324415},
  
  author = {Allen, Benjamin S. and Ezell, Matthew A. and Jacobsen, Douglas and Lueninghoener, Cory and Peltz, Paul and Roman, Eric and Wofford, J. Lowell},
  
  title = {Modernizing the HPC System Software Stack},
  
  publisher = {Zenodo},
  
  year = {2020},
  
  copyright = {Creative Commons Attribution 4.0 International}
}

@inproceedings{gamblin2015,
author = {Gamblin, Todd and LeGendre, Matthew and Collette, Michael R. and Lee, Gregory L. and Moody, Adam and de Supinski, Bronis R. and Futral, Scott},
title = {The Spack package manager: bringing order to HPC software chaos},
year = {2015},
isbn = {9781450337236},
publisher = {Association for Computing Machinery},
address = {New York, NY, USA},
url = {https://doi.org/10.1145/2807591.2807623},
doi = {10.1145/2807591.2807623},
booktitle = {Proceedings of the International Conference for High Performance Computing, Networking, Storage and Analysis},
articleno = {40},
numpages = {12},
location = {Austin, Texas},
series = {SC '15}
}

@misc{Grafana,
  author = {Grafana},
  title = {{Grafana}},
  howpublished = "\url{https://github.com/grafana/grafana}",
  year = {2017}, 
  note = "[Online; accessed 20-March-2026]"
}

@software{llview,
  author       = {Müller, Yannik and
                  Souza Mendes Guimarães, Filipe and
                  Karbach, Carsten and
                  Frings, Wolfgang},
  title        = {LLview},
  month        = may,
  year         = 2024,
  publisher    = {Zenodo},
  version      = {v2.3.0-base},
  doi          = {10.5281/zenodo.11236449},
  url          = {https://doi.org/10.5281/zenodo.11236449},
}

@InProceedings{reframe,
  author     = {Karakasis, Vasileios and Manitaras, Theofilos and Rusu, Victor Holanda and
                Sarmiento-P{\'e}rez, Rafael and Bignamini, Christopher and Kraushaar, Matthias and
                Jocksch, Andreas and Omlin, Samuel and Peretti-Pezzi, Guilherme and
                Augusto, Jo{\~a}o P. S. C. and Friesen, Brian and He, Yun and Gerhardt, Lisa and
                Cook, Brandon and You, Zhi-Qiang and Khuvis, Samuel and Tomko, Karen},
  title      = {Enabling Continuous Testing of {HPC} Systems Using {ReFrame}},
  booktitle  = {Tools and Techniques for High Performance Computing},
  editor     = {Juckeland, Guido and Chandrasekaran, Sunita},
  year       = {2020},
  month      = mar,
  series     = {Communications in Computer and Information Science},
  volume     = {1190},
  pages      = {49--68},
  address    = {Cham, Switzerland},
  publisher  = {Springer International Publishing},
  doi        = {10.1007/978-3-030-44728-1_3},
  venue      = {Denver, Colorado, USA},
  eventdate  = {2019-11-17/2019-11-18},
  eventtitle = {{HUST} - Annual Workshop on {HPC} User Support Tools},
  isbn       = {978-3-030-44728-1},
  issn       = {1865-0937},
}

@software{JUBE,
  author       = {Breuer, Thomas and
                  Wellmann, Julia and
                  Souza Mendes Guimarães, Filipe and
                  Himmels, Carina and
                  Luehrs, Sebastian},
  title        = {JUBE},
  month        = nov,
  year         = 2023,
  publisher    = {Zenodo},
  version      = {REL-2.6.1},
  doi          = {10.5281/zenodo.10228432},
  url          = {https://doi.org/10.5281/zenodo.10228432},
}

@InProceedings{buildtest,
author="Siddiqui, Shahzeb",
editor="Juckeland, Guido
and Chandrasekaran, Sunita",
title="Buildtest: A Software Testing Framework with Module Operations for HPC Systems",
booktitle="Tools and Techniques for High Performance Computing",
year="2020",
publisher="Springer International Publishing",
address="Cham",
pages="3--27",
isbn="978-3-030-44728-1"
}

@misc{OLCF,
title = {OLCF Test Harness},
author = {Dietz, Dan and Hagerty, Nick and Berrill, Mark and Brim, Michael and Budiardja, Reuben and Joubert, Wayne and Melesse Vergara, Veronica and Tharrington, Arnold and Fillers, Thomas},
abstractNote = {Acceptance and regression testing of a High Performance Computing (HPC) system requires an automated and reproducible framework and tool for running and logging results. Manually running tests across a system is labor intensive and prone to reproducibility errors. The OLCF Test Harness (OTH) provides a framework in which to document required tests for a HPC system. The OTH then provides tools to execute and log results of these tests in an automated fashion.},
doi = {10.11578/dc.20250328.1},
url = {https://doi.org/10.11578/dc.20250328.1},
howpublished = {[Computer Software] \url{https://doi.org/10.11578/dc.20250328.1}},
year = {2025},
month = {jul}
}

@misc{carpenter2022,
  author       = {Carpenter, Paul and
                  Utz, Uwe-Haus and
                  Narasimhamurthy, Sai and
                  Suarez, Estela},
  title        = {Heterogeneous  High Performance Computing},
  month        = feb,
  year         = 2022,
  publisher    = {Zenodo},
  doi          = {10.5281/zenodo.6090425},
  url          = {https://doi.org/10.5281/zenodo.6090425},
}

@article{kogge2008,
author = {Kogge, Peter and Borkar, S. and Campbell, Dan and Carlson, William and Dally, William and Denneau, Monty and Franzon, Paul and Harrod, William and Hiller, Jon and Keckler, Stephen and Klein, Dean and Lucas, Robert},
year = {2008},
month = {01},
pages = {},
title = {ExaScale Computing Study: Technology Challenges in Achieving Exascale Systems},
volume = {15},
journal = {Defense Advanced Research Projects Agency Information Processing Techniques Office (DARPA IPTO), Techinal Representative}
}

@article{babelstream2018,
title = {Evaluating attainable memory bandwidth of parallel programming models via BabelStream},
year = {2018},
issue_date = {January 2018},
publisher = {Inderscience Publishers},
address = {Geneva 15, CHE},
volume = {17},
number = {3},
issn = {1742-7185},
journal = {Int. J. Comput. Sci. Eng.},
month = jan,
pages = {247–262},
numpages = {16}
}

@article{heldens2021,
author = {Heldens, Stijn and Hijma, Pieter and Werkhoven, Ben Van and Maassen, Jason and Belloum, Adam S. Z. and Van Nieuwpoort, Rob V.},
title = {The Landscape of Exascale Research: A Data-Driven Literature Analysis},
year = {2020},
issue_date = {March 2021},
publisher = {Association for Computing Machinery},
address = {New York, NY, USA},
volume = {53},
number = {2},
issn = {0360-0300},
url = {https://doi.org/10.1145/3372390},
doi = {10.1145/3372390},
journal = {ACM Comput. Surv.},
month = mar,
articleno = {23},
numpages = {43}
}

@InProceedings{logmap2011,
author="Jim{\'e}nez-Ruiz, Ernesto
and Cuenca Grau, Bernardo",
editor="Aroyo, Lora
and Welty, Chris
and Alani, Harith
and Taylor, Jamie
and Bernstein, Abraham
and Kagal, Lalana
and Noy, Natasha
and Blomqvist, Eva",
title="LogMap: Logic-Based and Scalable Ontology Matching",
booktitle="The Semantic Web -- ISWC 2011",
year="2011",
publisher="Springer Berlin Heidelberg",
address="Berlin, Heidelberg",
pages="273--288",
isbn="978-3-642-25073-6"
}

@inproceedings{shamis2015ucx,
  title={UCX: an open source framework for HPC network APIs and beyond},
  author={Shamis, Pavel and Venkata, Manjunath Gorentla and Lopez, M Graham and Baker, Matthew B and Hernandez, Oscar and Itigin, Yossi and Dubman, Mike and Shainer, Gilad and Graham, Richard L and Liss, Liran and others},
  booktitle={2015 IEEE 23rd Annual Symposium on High-Performance Interconnects},
  pages={40--43},
  year={2015},
  organization={IEEE}
}

@inproceedings{herten24,
author = {Herten, Andreas and Achilles, Sebastian and Alvarez, Damian and Badwaik, Jayesh and Behle, Eric and Bode, Mathis and Breuer, Thomas and Caviedes-Voulli\`{e}me, Daniel and Cherti, Mehdi and Dabah, Adel and El Sayed, Salem and Frings, Wolfgang and Gonzalez-Nicolas, Ana and Gregory, Eric B. and Mood, Kaveh Haghighi and Hater, Thorsten and Jitsev, Jenia and John, Chelsea Maria and Meinke, Jan H. and Meyer, Catrin I. and Mezentsev, Pavel and Mirus, Jan-Oliver and Nassyr, Stepan and Penke, Carolin and R\"{o}mmer, Manoel and Sinha, Ujjwal and von St. Vieth, Benedikt and Stein, Olaf and Suarez, Estela and Willsch, Dennis and Zhukov, Ilya},
title = {Application-Driven Exascale: The JUPITER Benchmark Suite},
year = {2024},
isbn = {9798350352917},
publisher = {IEEE Press},
url = {https://doi.org/10.1109/SC41406.2024.00038},
doi = {10.1109/SC41406.2024.00038},
booktitle = {Proceedings of the International Conference for High Performance Computing, Networking, Storage, and Analysis},
articleno = {32},
numpages = {45},
location = {Atlanta, GA, USA},
series = {SC '24}
}

@article{ang2010graph500,
author = {Ang, James and Barrett, Brian and Wheeler, Kyle and Murphy, Richard},
year = {2010},
month = {01},
pages = {},
title = {Introducing the graph 500}
}

@misc{jpwr,
  author = {Juelich Supercomputing Centre},
  title = {{JSC power tool}},
  howpublished = "\url{https://github.com/FZJ-JSC/jpwr}",
  year = {2023}, 
  note = "[Online; accessed 20-March-2026]"
}

@misc{osu_bench,
  author = {Network-Based Computing Laboratory},
  title = {{OSU Micro-Benchmarks}},
  howpublished = "\url{http://
mvapich.cse.ohio-state.edu/benchmarks/}",
  year = {2023}, 
  note = "[Online; accessed 20-March-2026]"
}

@misc{Hubcast,
  author = {LLNL},
  title = {{Hubcast}},
  howpublished = "\url{https://github.com/LLNL/hubcast}",
  year = {2023}, 
  note = "[Online; accessed 20-March-2026]"
}

@misc{Pavilion2,
  author = {LANL},
  title = {{Pavilion2}},
  howpublished = "\url{https://github.com/hpc/pavilion2}",
  year = {2017}, 
  note = "[Online; accessed 20-March-2026]"
}

@misc{BenchPro,
  author = {TACC},
  title = {{BenchPro}},
  howpublished = "\url{https://github.com/TACC/benchpro}",
  year = {2023}, 
  note = "[Online; accessed 20-March-2026]"
}

\end{document}